\documentclass[aps,twocolumn,showpacs,superscriptaddress,letterpaper]{revtex4}
\usepackage[dvips]{graphicx}
\usepackage{amsmath} 
\usepackage{amssymb}
\usepackage{dcolumn}
\usepackage{pifont}

\begin{document}
\title{Thermal diffusion by Brownian motion induced fluid stress}

\author{Jennifer Kreft}
\affiliation{Institute of Physics, Academia Sinica, Taipei, Taiwan}
\author{Yeng-Long Chen}
\affiliation{Institute of Physics, Academia Sinica, Taipei, Taiwan}
\affiliation{Research Center for Applied Science, Academia Sinica, Taipei, Taiwan}
\date{\today}
\begin{abstract}
The Ludwig-Soret effect, the migration of a species due to a temperature 
gradient, has been extensively studied without a complete picture of its 
cause emerging. Here we investigate the dynamics of DNA and spherical particles subjected to a thermal gradient using a combination of 
Brownian dynamics and the lattice Boltzmann method.  We observe that the DNA 
molecules will migrate to colder regions of the channel, an observation also 
made in the experiments of Duhr, et al\cite{braun}. In fact, the thermal diffusion 
coefficient found agrees quantitatively with the experimental value. 
We also observe that the thermal diffusion coefficient decreases as the radius of the studied spherical particles increases.
Furthermore, we observe that the thermal fluctuations-fluid momentum flux coupling induces a gradient in the stress which leads to thermal migration in both systems.

\end{abstract}

\pacs{}

\maketitle

\section{Introduction}
The first observations of the migration of a species due to a temperature gradient were reported by Ludwig and Soret more than 100 years ago \cite{ludwig,soret1}.
Since that time, the effect has been observed in many multicomponent systems including fluid mixtures, colloidal suspensions, and polymer solutions \cite{wiegand}.
The mass flux of a species is described by Fick's law with an added term to account for thermal diffusion:
\begin{equation}
J_y=-\rho D\frac{dc}{dy}-\rho D_Tc(1-c)\frac{dT}{dy}
\label{fick}
\end{equation}
where $J_y$ is the particle flux in the y-direction.  The first term denotes diffusion due to a density gradient: $D$ is the molecular diffusion coefficient, $c$ is the mass fraction of the migrating species, and $\rho$ is the mass density.
The second term describes diffusion due to the temperature gradient: $D_T$ is the thermal diffusion coefficient and $T$ is the temperature.
At steady state, $J_y=0$ and the Soret coefficient, $S_T$ is defined:
\begin{equation}
S_T\equiv \frac{D_T}{D}=-\frac{1}{c(1-c)}\frac{\partial c/\partial y}{\partial T/\partial y}.
\end{equation}
Note that $S_T$ can be positive or negative depending on whether the species migrates to the hot ($S_T < 0$) or cold ($S_T >0$) region.

In general, the thermal diffusion coefficient for a molecule will depend on many factors\cite{wiegand}.
For example, a recent simulation conducted on DNA tightly confined to nanometer scale channels showed migration towards the heated region\cite{khare}.
In contrast, experiments on DNA unconfined in the direction of the temperature gradient show the polymer migrating to the cold region\cite{braun}.
Other factors shown in experiment to be important include the average temperature of the system, the solvent used, and electrostatic effects \cite{protein, peo, miscelles}.

Since many factors can play an important role in determining the thermal diffusion coefficient, theoretical predictions and experimental data sometimes do not agree.
The experiments in \cite{shiunduh} used colloidal particles of many different materials and showed that $D_T$ will increase with increasing particle diameter, $a$, when some aqueous solutions are used to suspend the colloids and decrease with increasing $a$ for others.
These experiments suggest that the surface interactions between the particle and the solvent play important roles in the particles' thermodiffusion coefficient.
Another experiment using carboxyl modified polystyrene particles showed that $D_T$ increases with increasing $a$.
The authors propose a model based on local equilibrium conditions that predicts $D_T$ will only increase with increasing $a$\cite{brauntheory}.
They observe this trend when the magnitude of the temperature gradient is much smaller than that in the experiments reported in \cite{shiunduh}.
They therefore surmise that the difference between their results and those in \cite{shiunduh} are due to differences in gradient magnitude and nonlinear effects in large gradients.
Another model based on volume transport theory has proposed that $D_T$ for dilute solutions of large molecules depends only on the solvent isobaric thermal expansion coefficient assuming that the pressure is uniform throughout the fluid\cite{brenner05,brenner}.

Several theoretical studies have proposed that the thermal diffusion of colloids or polymers is a surface driven phenomenon.  This approach was first adopted by Ruckenstein \cite{ruckenstein}. Piazza and Guarino use this model to qualitatively predict the role of electrostatics in the thermal diffusion of charged micelles \cite{miscelles}.
In general, these studies use the hydrodynamic equations to calculate a pressure gradient and/or volume force acting locally on the particles that is induced by a non-uniform distribution of solvent molecules or temperature dependent solvent-particle interaction\cite{miscelles, ruckenstein,schimpf04,piazza04,morozov1,morozov2}.
Others also include a macroscopic pressure gradient due to the response of the solvent alone to the temperature gradient and this model nearly quantitatively reproduces the mobility of polymers as measured in experiments\cite{schimpf04}.
Another approach begins with the kinetic theory of diffusion in a nonuniform temperature field to derive expressions for the Soret coefficient which allow for both positive and negative values of $S_T$ \cite{bandb}.

Here, we present results from a lattice Boltzmann simulation of $\lambda$ DNA in a microchannel subjected to a thermal gradient that quantitatively agrees with published experimental values for the thermal diffusion coefficient.
We show that a  \emph{non-equilibrium} stress develops from particle fluctuations.
This component of the solvent characteristics causes thermal migration of the species.
We also investigate the thermal diffusion of small, spherical particles and show that $D_T$ decreases with increasing diameter, independent of the magnitude of the thermal gradient.

\section{Lattice Boltzmann with Brownian Dynamics Simulation.}
We use a simulation based on the lattice Boltzmann method (LBM) for the fluid coupled with a worm-like chain (WLC) model with Brownian dynamics (BD) for the polymer\cite{ahlrichs:1999,ahlrichs:1998,usta:2005,usta:2006}.
The fundamental quantity in the LBM is the fluid velocity distribution function, $n_i(\mathbf{r},t)$, which describes the fraction of fluid particles with a discretized velocity, $\mathbf{c_i}$, at each lattice site\cite{ladd:2001,ladd:1994}.
To describe the velocities at each node, a 19 discrete velocity scheme in three dimensions is used.
The velocities can be represented by: $(0, 0, 0)$, $(\pm 1, 0, 0)$, $(0, \pm 1,0)$, $(0, 0, \pm 1)$, $(\pm 1, \pm 1, 0)$, $(0, \pm 1, \pm 1)$,  and $(\pm 1, 0, \pm 1)$ and have magnitudes $c_i=|\mathbf{c_i}|= 0, 1,$ or $\sqrt{2}$.
The maximum velocity in the simulation is given by the speed of sound: $c_s=\sqrt{1/3} \Delta x/ \Delta \tau$ where $\Delta x$ is the lattice spacing and $\Delta \tau$ is the time step.
The distributions will be Maxwell-Boltzmann at equilibrium and can be represented by a second order expansion:
\begin{equation}
n_i^{eq}=\rho a^{c_i} [1 +(\mathbf{c_i} \cdot \mathbf{u})/c_s^2 + \mathbf{uu}:(\mathbf{c_ic_i}-c_s^2\mathbf{I})/(2c_s^4)]
\end{equation}
where $\mathbf{u}$ is the local velocity.  The coefficients $a^{ci}$ are found by satisfying the isotropy condition:
\begin{equation}
\sum_i a^{ci}\mathbf{c}_{i\alpha}\mathbf{c}_{i\beta}\mathbf{c}_{i\gamma}\mathbf{c}_{i\delta}=c_s^4(\delta_{\alpha\beta}\delta_{\gamma\delta}+\delta_{\alpha\gamma}\delta_{\beta\delta}+\delta_{\alpha\delta}\delta_{\beta\gamma})
\end{equation}
where $\alpha, \beta, \gamma,$ and $\delta$ represent the $x$, $y$, or $z$ axis.  The equilibrium conditions for the density $\rho$, momentum density $\mathbf{j}$, and momentum flux density $\mathbf{\Pi}$:
\begin{equation}
\rho=\sum_i n_i^{eq}
\end{equation}
\begin{equation}    
\mathbf{j}=\rho\mathbf{u}=\sum_i\mathbf{c_i}\cdot n_i^{eq}     
\end{equation}
\begin{equation}
\mathbf{\Pi}=\rho (\mathbf{uu} +c_s^2\mathbf{I})=\sum_i n_i^{eq} \cdot \mathbf{c_ic_i}.  
\end{equation}
must also be satisfied.

At each time step, the fluid particles will collide with their nearest and next nearest neighbors.  The velocity distributions will evolve according to:
\begin{equation}
  n_i(\mathbf{r}+\mathbf{c_i}\Delta\tau, t+\Delta\tau)=n_i(\mathbf{r},t)+\mathbf{L}_{ij}[n_j(\mathbf{r},t)-n_j^{eq}(\mathbf{r},t)]
\end{equation}
where $\mathbf{L}$ is a collision operator for fluid particle collisions such that the fluid always relaxes towards the equilibrium distribution. 
In the limit of small Knudson and Mach number, this equation has been shown to be equivalent to the Navier Stokes equation \cite{benzi:1992}.

Collisions are simplified by transforming the $n_i$ from velocity space into the hydrodynamic moments, $M_q=\mathbf{m} \cdot \mathbf{n}$, where $M_q$ is the $q^{th}$ moment of the distribution, $\mathbf{m}$ is the transformation matrix, and $\mathbf{n}=(n_0,n_1,...n_{18})$.  The density, momentum density, momentum flux, and the kinetic energy flux constitute the nineteen moments; however, kinetic energy flux moments conserve energy and are considered 'ghost' moments.

The collision operator is chosen to be a diagonal matrix with elements $\tau_0^{-1},\tau_1^{-1},...,\tau_{18}^{-1}$ where $\tau_q$ is the characteristic relaxation time of the moment $q$.  
The conserved moments, density and momentum, are considered to have an infinite relaxation time and thus $\tau_{0,1,2,3}^{-1}=0$.  
The other moments have a single relaxation time, $\tau_s$, as in the Bhatanagar-Gross-Krook (BGK) model \cite{BGK}.
The relaxation time is constrained to be smaller than the fluid momentum diffusion time across the system, $\tau_s < \rho H^2/\eta$ where the shear viscosity, $\eta$ is given by: $\eta=\rho c_s^2 (\tau_s-0.5)$ for $\tau_s>0.5$ \cite{benzi:1992}.
In our simulations, we have $\tau_s=1.0$.

A worm-like chain model is adopted for the $\lambda$ DNA in the simulation.
The model has been parameterized to capture molecular dynamics in bulk solution at $T=298K$, and has been used to accurately predict the diffusivity of 48.5 kbps YOYO-stained DNA in microchannels \cite {wisc1,wisc2,wisc3}.

Each molecule is represented by 11 beads connected by 10 worm-like springs.
The position and velocity of the beads are updated using the explicit Euler method.
\begin{equation}
\mathbf{u}(t + \Delta t)=\mathbf{u}(t)+ \mathbf{f}(t)\Delta t/m
\end{equation}
\begin{equation}
\mathbf{x}(t + \Delta t)=\mathbf{x}(t) + \mathbf{u}(t)\Delta t
\end{equation}
where $\mathbf{f}(t)$ is the total force acting on the bead, $\mathbf{u}(t)$ is the velocity, and $\mathbf{x}(t)$ is the position of the bead with mass $m$ at time $t$.  The time step for the beads is $\Delta t$.  The forces acting on the bead include excluded volume effects, the elastic force of the springs, hydrodynamic interactions with the solvent, and the Brownian motion of the particles.

The excluded volume interactions between segments are calculated using a Gaussian excluded volume potential that leads to self-avoiding walk statistics:
\begin{equation}
U_{ij}^{ev}=\frac{1}{2}k_BT \nu N_{ks}^2(\frac{3}{4 \pi S_s^2})\exp{(\frac{-3\lvert \mathbf{r}_i-\mathbf{r}_j \rvert^2}{4S_s^2})}
\end{equation}
where $\nu =\sigma_k^3$ is the excluded volume interaction parameter, $N_{ks}=19.8$ is the number of Kuhn segments per spring, and $S_s^2=(N_{ks}/6)\sigma_k^2$ is the characteristic size of the coarse grained beads.

An experimentally determined force-extension relation is used to calculate the elastic force on a bead\cite{bustamante}:
\begin{equation}
\mathbf{f^s_{ij}}=\frac{k_BT}{2\sigma_k}[(1-\frac{|\mathbf{r_j}-\mathbf{r_i}|}{N_{ks}\sigma_k})^2 +4\frac{|\mathbf{r_j}-\mathbf{r_i}|}{N_{ks}\sigma_k}-1]\frac{\mathbf{r_j}-\mathbf{r_i}}{|\mathbf{r_j}-\mathbf{r_i}|}
\end{equation}
This was found when measuring the force-extension relation for the entire chain.
The force-extension relation is accurate when $N_{ks} \gg 1$ and we apply this equation to single chain segments of $N_{ks}=19.8$.

The fluid exerts a frictional force on the beads, given by:
\begin{equation}
\mathbf{F}_f=-\zeta(\mathbf{u}_p-\mathbf{u}_f)
\end{equation}
where $\mathbf{u}_p$ is the velocity of the bead, $\mathbf{u}_f$ is the velocity of the fluid at the bead position, $\zeta=6\pi\eta a$ is the friction coefficient, and $a$ is the hydrodynamic radius of the bead \cite{ahlrichs:1998}.  The simulation lattice size, $\Delta x$, is chosen to be 0.5 $\mu m$.  For our model DNA chain, each bead has a hydrodynamic radius of $a=0.077 \mu m$, or 0.154$\Delta x$ \cite{chen,wisc2}.   Since the beads' positions are not limited to the lattice where the fluid velocity is well defined, $\mathbf{u}_f$ at the position of the bead is determined by linearly interpolating the velocities of the nearest neighbor lattice sites such that $\mathbf{u}_f=\sum_{i\in n.n.}w_i\mathbf{u}_i$.  The weighting factors $w_i$ are normalized and $\mathbf{u}_i$ represents the fluid velocity at site $i$. 
The momentum transfer to the bead is $\Delta \mathbf{j}= -\mathbf{F}_f\Delta t/\Delta x^3$.
The bead will also transfer this momentum to the fluid.  The momentum transfer from the bead to a neighbor site $i$ with velocity $q$ is $\Delta \mathbf{f}_i=w_i\rho a_{c_q}\Delta \mathbf{j} \cdot \mathbf{c}_q$ \cite{ahlrichs:1998}.

The beads also undergo Brownian motion.   The thermal fluctuations of the beads will be drawn from a Gaussian distribution with zero mean and a variance that varies with the bead height: $\sigma_v=2k_BT(y)\zeta \Delta t$.  Here, 
\begin{equation}
T(y)=\frac{2(T_{hot}-T_{cold})}{Y_{max}} \vert(Y_{max}/2-y)\vert +T_{cold}
\end{equation}
where $Y_{max}$ is the width of the channel, $y$ is the position of the bead in the channel, $T_{hot}$ is the maximum and $T_{cold}$ is the minimum temperature in the channel.  This leads to a saw tooth shape of the temperature plotted as a function of $y$.  It should be noted that the temperature gradient in the system is only implemented here; all other forces and fluid properties are independent of location.

For this work, 50 polymers were simulated in a container of size 20 $\mu m$ x 20 $\mu m$ x 20 $\mu m$.  Periodic boundary conditions were imposed in all directions unless otherwise noted.
The time step for the fluid is $\Delta \tau =8.8 \times 10^{-5}$s, and for the polymer $\Delta t = 1.72 \times 10^{-5}$s as calculated using $T=T_{cold}$.
The total simulation time was 1760 seconds.
Data was recorded once every 17.6 seconds; the final 40 time steps were used to determine $D_T$.

\section{Results}

\subsection {Thermal migration of $\lambda$ DNA}
\begin{figure} [hb]\centering

  \includegraphics[width=6cm, angle=270]{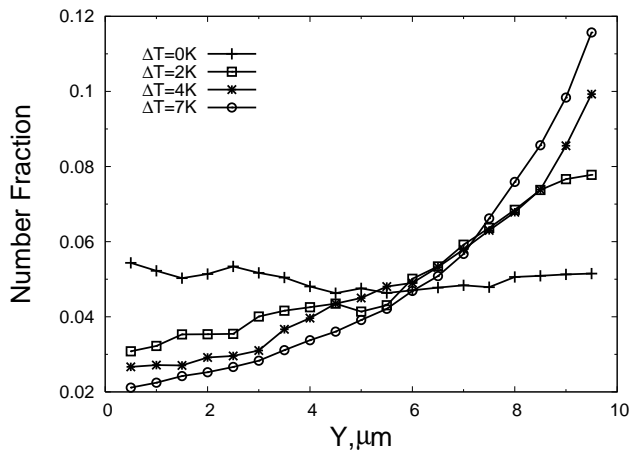} \caption{Number fraction, $n/n_{total}$ where $n$ is the number of beads whose center of mass is between $y-0.25$ and $y+0.25$ and $n_{total}$ is the total number of beads, as a function of height for $\lambda$-DNA subjected to different temperature gradients.  Data are averaged over the final 40 time steps of five simulations started from different random initial conditions. Shown is the average of the two mirror-image halves of the periodic system.}   
\label{fig1}
\end{figure}
As shown in Figure \ref{fig1}, the DNA accumulates in the center of the channel where the temperature minimum is found.  The simulations with a larger temperature gradient result in a larger concentration gradient.  The profile is nearly flat when the temperature is uniform in the channel.
These results are in qualitative agreement with the work of Duhr and Braun \cite{braun}.  

For quantitative comparison, we use the same equation found in \cite{braun}:
\begin{equation}
\frac{c(z)}{c_0}=exp\left[-S_T(T(z)-T_0)\right]
\label{czeq}
\end{equation}
where $c(z)$ is the concentration of DNA at position z, $c_0$ is the maximum concentration, $T(z)$ is the temperature at z, $T_0$ is the temperature at the position of $c_0$, and $S_T$ is the Soret coefficient.
This equation is derived from Eqn. (\ref{fick}) with $J_y=0$ and $c \ll 1$.
The average Soret coefficient was found by fitting the density profile to Eqn. (\ref{czeq}) and solving for $S_T$.
As in \cite{braun}, we use $D=1.0 \mu m^2/s$ for $\lambda$ DNA to calculate $D_T$ from $S_T$ \footnote{To compare with the experimental results quantitatively, we use the same value for the molecular diffusion coefficient of $\lambda$ DNA.  However, an experimentally determined value of $D=0.48 \mu m^2/s$ was reported by Smith, Perkins and Chu in \cite{chu}.  Using this value would decrease $D_T$ by a factor of 0.48 as $D_T$ depends linearly on $D$.}.
We find, for the data presented in Figure \ref{fig1}, $D_T=0.38 \pm $0.1 $\mu m^2/sK$.
This is in good agreement with the value of 0.4 $\mu m^2/sK$ reported in \cite{braun}.

In \cite{braun}, identical thermal diffusion coefficients were measured for 27bp and 48.5 Kbp DNA.  
We find similar values of $D_T$  for  48.5 Kbp ($D_T=0.40 \pm $0.06 $\mu m^2/sK$), 19.4 Kbp ($D_T =0.46 \pm $0.06$ \mu m^2/sK$), and 67.9 Kbp DNA ($D_T=0.40 \pm $0.06 $\mu m^2/sK$) for $\Delta T=4K$.
The diffusion coefficient $D$ of the individual molecules were calculated according to:
\begin{equation}
D_L=D_{\lambda}(\frac{L}{L_{\lambda}})^{-0.588}
\end{equation}
where $D_{\lambda}$ is 1 $\mu m^2/s$ \footnotemark[\value{footnote}], $L$ is the length of the DNA, $L_{\lambda}$ is 48.5 Kbp, the length of $\lambda$ DNA, and $D_L$ is the molecular diffusion coefficient for DNA of length $L$\cite{wisc2}.

\begin{figure} [hb]\centering

  \includegraphics[width=6cm, angle=270]{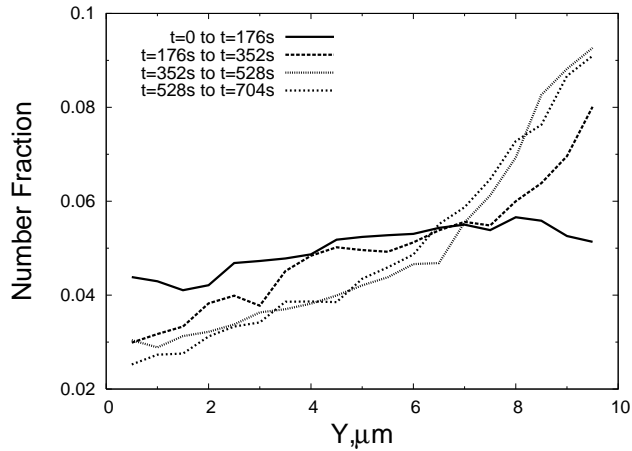} \caption{Number fraction, $n/n_{total}$ where $n$ is the number of beads whose center of mass is between $y-0.25$ and $y+0.25$ and $n_{total}$ is the total number of beads, as a function of height at different times for $\lambda$-DNA with $\Delta T=4K$.  Data are averaged over 10 time steps of five simulations started from different random initial conditions. Shown is the average of the two mirror-image halves of the periodic system.}   
\label{fig2}
\end{figure}
The development of the density profile over time can be seen in Fig. ~\ref{fig2}.  
The profile develops slowly and only after more than 300 seconds does the system reach steady state.
This time frame is similar to that observed by Duhr, Arduini, and Braun who report that several hundred seconds are needed to reach the steady state \cite{braun}.
\subsection {Mechanism of thermal migration}
\begin{figure} [hb]\centering

\includegraphics[width=6cm, angle=270]{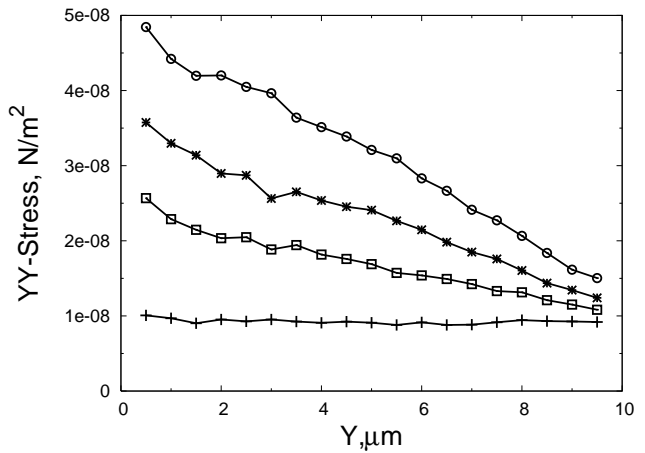} \caption{Stress in the y-direction across the channel for $\lambda$-DNA.  Data are averaged over the final 40 time steps of five simulations started from different random initial conditions. Shown is the average of the two mirror-image halves of the periodic system.}

\label{fig3}
\end{figure}

To understand why the polymers migrate to the colder regions, we investigated the dynamics of the solvent.  
Since the properties of the fluid were kept constant across the channel, the migration must result from interactions between the fluid and the polymers.  
Thus the momentum flux of the fluid within 2 lattice sites of a bead was recorded.
This quantity is known to be coupled to the DNA fluctuations \cite{ahlrichs:1998} and will contribute to the local fluid stress\cite{chapman}.
A gradient in $\Pi_{yy}=\sum_i n_i^{eq} \cdot \mathbf{c_ic_i}$ is observed for non-zero temperature gradient but is absent for simulations with uniform temperature as seen in Figure \ref {fig3}.
As gravity is absent, the stress is only due to the momentum flux \cite{chapman}.
This gradient in flux is therefore a gradient in stress which causes the polymers to migrate into the cold regions.

The difference in stress is due solely to the interaction of the beads and the fluid.  
Without the presence of the DNA, the fluid will relax back to equilibrium conditions and therefore uniform stress across the channel.
In fact, when the force the polymers exert on the fluid is set to zero in the simulation, no thermal diffusion is observed.
Thus the fluctuations of the polymers induce a local, short-lived gradient in stress which causes the thermal migration of the species.

The steady state is reached when the particle flux from the cold region to the hot region equals the flux in the reverse direction.
The particles will migrate only because of Brownian motion in this simulation.
Thus there is a well defined probability of a particle in the cold side receiving a sufficient Brownian kick to move to a higher stress region.
In the steady state, this probability times the number of beads in the cold side must equal the corresponding probability a particle will move towards the low stress area times the number of beads in the hot region.
Since the Brownian force depends on the viscosity of the fluid and the hydrodynamic radius of the particle, the thermal diffusion coefficient should also depend on these parameters.  We will investigate these dependencies within the limits of our simulation.

\subsection{Particle size effects}
\begin{figure} [hb]\centering

\includegraphics[width=6cm, angle=270]{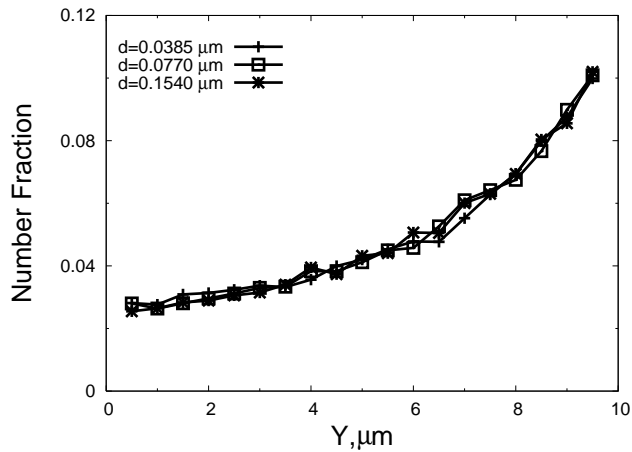} \caption{Number fraction, $n/n_{total}$ where $n$ is the number of beads whose center of mass is between $y-0.25$ and $y+0.25$ and $n_{total}$ is the total number of beads, as a function of height for spherical particles of different diameter with $\Delta T=4K$.  Data are averaged over the final 40 time steps of five simulations started from different random initial conditions. Shown is the average of the two mirror-image halves of the periodic system.}

\label{fig4}
\end{figure}
\begin{figure} [hb]\centering

\includegraphics[width=6cm, angle=270]{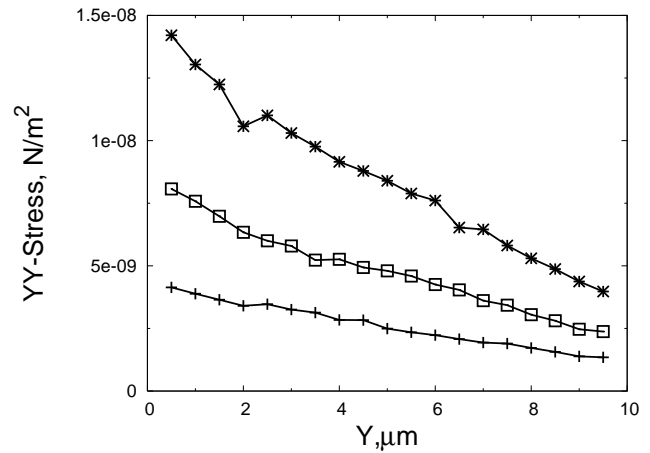} \caption{Stress in the y-direction across the channel for spherical particles of different diameter.  Data are averaged over the final 40 time steps of five simulations started from different random initial conditions. Shown is the average of the two mirror-image halves of the periodic system.}

\label{fig5}
\end{figure}

The polymer and fluid are coupled through the Brownian and viscous forces.  
Both terms depend on $\zeta=6\pi \eta a$; the viscous force explicitly and the Brownian term through the standard deviation of the distribution of the fluctuations.  
In the simulation, neither the hydrodynamic radius, $a$, or viscosity of the fluid, $\eta$, appears singly.
Thus, doubling the value of one term is analogous to doubling the other.
However, we consider changes to $\zeta$ to be changes to $a$ since experimental work has been conducted on the effects of changing colloidal particle diameter rather than changing the viscosity of the fluid \cite{brauntheory, shiunduh}.

The simulations were conducted using 100 individual spheres not attached to each other by springs. 
We investigated particles with a hydrodynamic radius of 0.0385, 0.077, 0.154, and 0.231 $\mu m$ with a temperature difference of $4K$ across the 10$\mu m$ channel.  
The diffusion coefficient, $D$, of each species was calculated according to $D=k_BT_{cold}/6\pi \eta a$ where $a$ is the hydrodynamic radius of the particle and $\eta$ is the viscosity of the solution.
Similar values were obtained for the Soret coefficient, $S_T$, as can be seen in Figure \ref{fig4}.
Here, the density profiles are nearly identical for all diameters.
However, because of the different diffusion coefficients, $D$, larger values for $D_T$ were obtained for the smaller particles.  See Table ~\ref{table1} for a comparison of values.
It has been suggested that $D_T$ decreases with increasing $a$ because the gradient is too steep to allow the particles to reach local equilibrium.
We therefore decreased the temperature difference to $2K$ to test this hypothesis. 
However, as Table \ref{table1} shows, the values of $D_T$ do not change appreciably.
This suggests the decrease of $D_T$ with increasing radius is not due to a particles being farther out of local equilibrium.
Indeed, all of the above simulations meet the criteria,  $(aS_T)^{-1} > \nabla T$,  as set by the local equilibrium condition in \cite{brauntheory}.
\begin{table}
\caption{Values of the thermal diffusion coefficient, $D_T$, given in $\mu m^2/sK$ for spheres of different hydrodynamic radius, $a$, and temperature difference, $\Delta T$ across the 10 $\mu m$ channel.}
\label{table1}
\begin{tabular}{|c|c|c|c|}
\hline
$\Delta T$&$a=0.0385 \mu m$&$a=0.077 \mu m$&$a=0.154 \mu m$\\
\hline
\hline
4K&$2.1 \pm 0.3$&$1.12 \pm 0.06$&$0.59 \pm 0.04$\\
\hline
2K&$2.3 \pm 0.4$&$1.1 \pm 0.2$&$0.60 \pm 0.01$\\
\hline
\end{tabular}
\end{table}

We suggest a different explanation based on the analysis of the non-equilibrium stress induced by the thermal gradient.
The fluid stress gradient can be seen in Figure ~\ref{fig5}.
The larger diameter particles induce a steeper gradient in the stress than do the smaller spheres.
However, the steady state density profile is the same for all particles due to the differences in the Brownian force on the particles.
Since the standard deviation of the distribution is proportional to the hydrodynamic radius, larger particles experience larger fluctuations.
This is exactly balanced by the steeper gradient induced, leading to steady state density profiles that are nearly the same for all particle sizes.

\section{Conclusions.}
We use a lattice-Boltzmann simulation to investigate the mechanism behind thermal diffusion of $\lambda$-DNA and spherical particles.
This method allows us to capture the non-equilibrium stress in the fluid due to the temperature gradient.
We find good agreement with experimental results for the thermal diffusion coefficient, $D_T$, for $\lambda$-DNA \cite{braun}.
It is also observed that $D_T$ decreases as the diameter, $a$, of diffusing spherical particles increases in partial agreement with experiments by Shiunduh, \emph{et al.} \cite{shiunduh}.

The thermal diffusion coefficient is observed to decrease if the size of the diffusing species is increased.
In our work, unlike in experiments, the fluid characteristics such as viscosity are held constant across the channel.
It is also noted that decreasing the temperature difference across the channel did not change these results.
Therefore the dependence on particle size can not be explained as a result of the spheres not reaching local equilibrium with the fluid or the fluid characteristics varying too much across the channel.

Instead, the non-equilibrium component of the fluid flow is observed to be linked with the migration of the solute.
The induced stress is significantly more in the hot region than in the cold.
Thus particles will migrate to the cold regions.
The thermal diffusion coefficient will therefore depend on factors influencing the interaction between the solvent and solute.

This picture is in agreement with the theoretical work of several authors \cite{ruckenstein,piazza04,morozov1,morozov2,schimpf04,miscelles} who investigate local pressure gradients induced by non-isotropic interactions between the solute and solvent.
However, in those studies, the specifics of the interaction of the diffusing particle and surrounding fluid played an important role in determining $S_T$.
Here, we have quantitatively predicted the thermal diffusion coefficient of DNA as well as the time scales of the phenomenon without including the details of the interaction between the polymer and solvent.
Instead of these characteristics being important, only the Brownian motion of the particle induces a local stress gradient in the fluid.
This may indicate that hydrodynamic memory effects, already shown to be important in the mass diffusion of polymers and colloids, should be considered when developing theories of thermal diffusion \cite{hinch,florin,colloidmemory,polymermemory}.
We therefore hope that this work will inspire new directions in theoretical examinations of thermal diffusion.

\bibliographystyle{apsrev}
\bibliography{soret,lbsims,nsf,labonchip}          

\end{document}